\documentclass[conference]{IEEEtran}
\IEEEoverridecommandlockouts
\usepackage{cite}
\usepackage{amsmath,amssymb,amsfonts}
\usepackage{algorithmic}
\usepackage{graphicx}
\usepackage{textcomp}
\usepackage{xcolor}
\usepackage{algorithm}
\usepackage{booktabs}
\usepackage{url}
\usepackage{hyperref}
\usepackage{booktabs}
\usepackage{siunitx}
\usepackage{multirow}
\usepackage{xcolor}
\usepackage{listings}
\usepackage{tabularx}
\usepackage{colortbl}  

\definecolor{added}{rgb}{0.0, 0.5, 0.0}  

\newcommand{\dataset}{\texttt{JamendoMaxCaps}}
\def\BibTeX{{\rm B\kern-.05em{\sc i\kern-.025em b}\kern-.08em
    T\kern-.1667em\lower.7ex\hbox{E}\kern-.125emX}}
\begin{document}

\title{JamendoMaxCaps: A Large Scale Music-caption Dataset with Imputed Metadata 
}


\author{
    \IEEEauthorblockN{Abhinaba Roy, Renhang Liu, Tongyu Lu, Dorien Herremans}
    \IEEEauthorblockA{Singapore University of Technology and Design} 
    \IEEEauthorblockA{\{abhnaba\_roy, renhang\_liu, tongyu\_lu, dorien\_herremans\}@sutd.edu.sg}
}

\maketitle
\begin{abstract}
We introduce \textbf{\dataset}, a large-scale music-caption dataset featuring over 362,000 freely licensed instrumental tracks from the renowned Jamendo platform. The dataset includes captions generated by a state-of-the-art captioning model, enhanced with imputed metadata. We also introduce a retrieval system that leverages both musical features and metadata to identify similar songs, which are then used to fill in missing metadata using a local large language model (LLLM). This approach allows us to provide a more comprehensive and informative dataset for researchers working on music-language understanding tasks. We validate this approach quantitatively with five different measurements. By making the \textbf{\dataset} dataset publicly available, we provide a high-quality resource to advance research in music-language understanding tasks such as music retrieval, multimodal representation learning, and generative music models.
\end{abstract}


\section{Introduction}
Music information retrieval is an active research field focused on developing computational techniques to analyze, organize, access, and even generate musical data. In recent years, rapid advances in deep learning and natural language processing have enabled novel approaches to music generation, where models can learn to generate music from textual descriptions~\cite{copet2023simple,melechovsky2023mustango,bhandari2024text2midi} or other modalities~\cite{kang2024video2music}. However, the lack of large-scale, high-quality datasets with rich musical and textual annotations has been a key challenge in driving progress in these music-language understanding tasks.

In the sound/speech domain, we find several expansive datasets such as LAION-Audio~\cite{wu2023large} and wav2vec~\cite{mei2024wavcaps}, each comprising over 500,000 audio-caption pairs. The music domain, however, is relatively limited in substantial publicly accessible resources. For instance, the widely utilized MusicCaps dataset \cite{2023musiclm} encompasses only approximately 5,000 examples. To address this scarcity, researchers frequently resort to data augmentation techniques to expand the dataset and generate synthetic captions based on the applied transformations \cite{melechovsky2023mustango}. Existing research has explored an approach that involves extracting musical features from pre-trained models and subsequently rephrasing them using a large language model, or even utilizing existing tags present in audio datasets in the form of metadata and using an LLM to generate captions based on them~\cite{2023lpmusiccaps}. However, extracting individual features from a song can be a time-consuming process, and in practice, metadata is often incomplete or unavailable for many music collections. 
\begin{figure}[t!]
    \centering
    \includegraphics[width=0.5\textwidth]{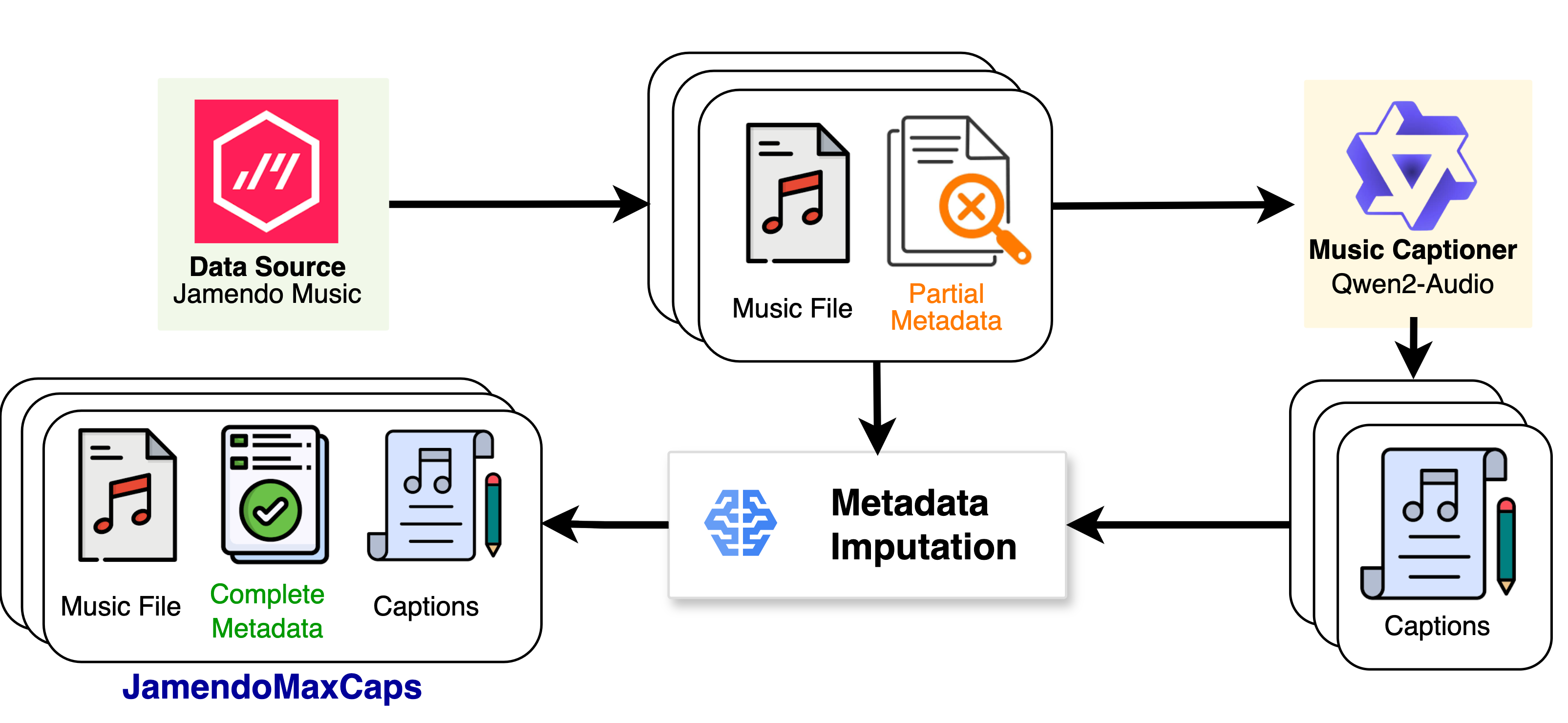}
    \caption{Overview of our proposed pipeline for the creation of the dataset.}
    \label{fig:overview}
\end{figure}

To mitigate these limitations, we introduce \textbf{\dataset}\footnote{Dataset: \url{https://huggingface.co/datasets/amaai-lab/JamendoMaxCaps}}
\footnote{Code: \url{https://github.com/AMAAI-Lab/JamendoMaxCaps}}, a large-scale music-caption dataset comprising over 362,000 freely licensed instrumental tracks from the renowned Jamendo platform, a source of user-contributed music. We leverage a state-of-the-art music captioning model~\cite{chu2024qwen2} to generate natural language descriptions for each track. To address the issue of incomplete metadata, we propose a multimodal metadata imputation approach that utilizes audio features as well a local large language model to expand the incomplete metadata. This allows us to populate fields such as genre, tempo, and variable tags like mood and instrumentation, providing a more comprehensive and informative dataset for researchers working on music-language understanding tasks.

We first construct a retrieval system that identifies similar songs based on the available musical and (partial) metadata features. We utilize the widely adopted MERT features to represent the musical content, and employ a sentence transformer to encode the partially available metadata. This query system combines both musical and metadata features, enabling efficient retrieval of similar songs based on their musical contents as well as associated (partial) metadata. We then use this retrieval system to retrieve the top 10 most similar songs with incomplete metadata in the dataset. We employ a local large language model to leverage in-context learning, where the model learns from the related examples to infer the missing metadata fields, such as genre, tempo, and variable tags like mood and instrumentation. The use of a local large language model is advantageous as it provides more control, flexibility, and cost-effectiveness compared to alternative approaches such as using a pre-trained GPT model hosted on a remote server. The selection of relevant in-context examples is critical, as the large language model can leverage these similar examples to infer better and predict the missing metadata fields. This contextual information is essential for the model to make well-informed predictions and populate the incomplete metadata in the dataset. We perform extensive evaluations to validate the choice of in-context examples as well as the quality of the imputed metadata, which is essential for the overall success and reliability of the dataset.

The key contributions of our work can be summarized as:
\begin{itemize}
    \item We present \textbf{\dataset}, a large-scale music-caption dataset featuring over 362,000 freely-licensed instrumental tracks from the Jamendo platform, along with generated captions.
    \item We also develop a metadata imputation approach that utilizes a local large language model (LLLM) to infer and populate missing metadata fields, such as genre, tempo, and variable tags like mood and instrumentation, resulting in a more comprehensive dataset.
    \item We construct a retrieval system that integrates musical and metadata features to identify similar songs, enabling efficient contextualization for the large language model to infer the missing metadata.
\end{itemize}

\section{Related Work}
In this section, we discuss recent text-music datasets, metadata imputation research as well as music feature retrieval systems. MusicCaps~\cite{2023musiclm}, one of the most popular text-music datasets, contains human-annotated captions for a limited set of around 5,000 music samples. Despite the high-quality descriptions, its small scale is not sufficient for data-hungry deep learning models. MusicBench~\cite{melechovsky2023mustango} further augments MusicCaps 11x and modifying the captions accordingly using an LLM. Still, it is limited to around 52k instances, each 10s in duration. The WavCaps dataset~\cite{2023wavcaps} expanded upon this concept by collecting over 400,000 audio-caption pairs from multiple online sources, employing heuristic filtering methods to ensure relevance. However, WaveCaps is more focused on generic audio rather than musical sounds. The recently released SongDescriber dataset~\cite{manco2023song}, contains 1.1k human-written text captions of 706 songs. The authors intended this dataset to be used as an evaluation set for text-to-music systems. SunoCaps~\cite{2024sunocaps} contains 256 prompt-generated music samples, showing an alternative way creating text-music dataset. Finally, Noise2music~\cite{huang2023noise2music} introduced the MuLaMCap dataset (MuLan-LaMDA Music Caption) which consists of 400k music-text pairs created by annotating AudioSet content with a data mining pipeline, but the dataset is not publicly available. 

The recently released Midicaps dataset~\cite{Melechovsky2024} contains over 178,000 music MIDI files with associated text captions. They used predefined feature extractors to get relevant features such as key, chords, and time signature, and used an LLM to generate captions. This dataset contains captions for MIDI music, however, not audio music files. The approach from  \cite{Melechovsky2024} aligns with existing methods that integrate symbolic representations with large language models for caption generation. In the waveform domain, LP-MusicCaps~\cite{2023lpmusiccaps} explored a tag-to-caption approach, where pre-existing metadata from audio libraries was used as input for a large language model, producing pseudo-captions to describe musical content. However, they relied on pre-existing metadata, which often exhibits notable incompleteness and inconsistencies in real-world scenarios. This limitation can hinder the comprehensive representation of musical content and potentially introduce biases in the generated captions.

Recent research has also explored metadata imputation to address incomplete or missing annotations in music datasets. For example, \cite{lee2018deep} proposed a collaborative filtering approach that combines content-based and user-generated metadata for automatic music tagging, leveraging deep learning to predict missing attributes based on musical embeddings. Similarly,~\cite{kim2024missing} introduced a transformer-based model that infers missing genre and mood labels from audio spectrograms, highlighting the role of multimodal learning in metadata completion. These works align with our use of large language models for metadata imputation, although our method uniquely incorporates a retrieval-based in-context learning strategy to enhance imputation accuracy.

In addition, music feature retrieval systems have played a crucial role in improving dataset utility. CLAP~\cite{2022clap} employed a contrastive learning approach for aligning music and text embeddings, enabling zero-shot music retrieval based on textual queries. MERT~\cite{2024mert} provided a powerful audio feature extractor designed specifically for music representation learning. Both models demonstrate the effectiveness of using pre-trained embeddings for understanding musical content. Salmonn~\cite{2024salmonn} is an integrated multiple-source audio features including Q-Former queries~\cite{2023blip}, Whisper features~\cite{2023whisper}, BEATs features~\cite{2023beats} to a generate music captions . Our approach builds upon these works by integrating retrieval-based metadata imputation with local large language models, aiming to ensure more contextually relevant annotations.

Despite the advancements made by these prior works, challenges remain in ensuring accurate, and scalable metadata imputation for large-scale music-caption datasets. The \textbf{\dataset} dataset aims to address these gaps by leveraging a hybrid approach that combines musical representation, retrieval-based contextualization, and in-context learning via locally hosted large language models.

\section{Method}
In this section, we detail our caption generation, retrieval system and metadata imputation approach for the \textbf{\dataset} dataset. Details regarding the dataset creation are in the next section.

\subsection{Caption Generation}
\label{sec:cap}
To generate high-quality textual descriptions for our large-scale dataset, we employ \textit{Qwen2-Audio}~\cite{chu2024qwen2}, a state-of-the-art music understanding model that has demonstrated competitive performance relative to other approaches (e.g., Salmonn~\cite{2024salmonn}). Our primary objective is to create informative captions focusing on instrumentation, genre, mood, rhythm, and potential listening scenarios as these descriptions serve as a crucial component of \dataset. To carry out caption generation we perform the following two steps:

\paragraph{Segmented Captioning for Temporal Variation} In order to capture potential temporal changes throughout each track, we segmented every song into consecutive 30-second clips, generating a caption for each segment. We discarded any final segment shorter than 15 seconds as we found that the caption tends to be of lower quality when the given tracks are too short. While multiple captions per track provide richer training data for music understanding and generation models, for subsequent retrieval and metadata imputation tasks, we focused on using only the first 30-second segment as the track’s representative excerpt. This decision balances computational efficiency with capturing core musical characteristics, such as initial instrumentation, tempo, and mood.

\paragraph{Efficient Inference and Prompt Design} To manage the computational demands of captioning over 362,000 tracks, we employed batched inference and leveraged 4-bit quantization to reduce model memory requirements, with minimal impact on caption quality. We set a maximum output length of 128 tokens per caption, ensuring that roughly 90\% of the generated text falls within this range, thus avoiding overly long or unwieldy outputs. Each 30-second clip was processed using a standardized prompt: \begin{quote} ``\textit{Generate a detailed caption for this music piece, primarily focusing on \textbf{instrumentation, genre, mood, rhythm, and scenario.}}'' \end{quote} Preliminary experiments indicate that explicitly mentioning both instrumentation and mood in the prompt yielded more coherent, contextually rich captions.

Our caption generation pipeline proves to be robust at scale, ultimately producing captions for more than 3 million 30-second audio segments, providing a reliable basis for subsequent metadata imputation and retrieval systems, as well as for broader music-language modeling efforts.

\subsection{Retrieval System}
To impute missing metadata fields, we first construct a retrieval system that identifies similar songs based on both musical and metadata features. For the musical content, we utilize MERT features~\cite{2024mert}~\footnote{We use \url{https://huggingface.co/m-a-p/MERT-v1-330M}} to represent the audio signals. We choose MERT due to its proven ability to understand as well as capture the \textit{musical} content in a song~\cite{2024mert}, whereas models like CLAP are more geared towards general audio sounds. 

To calculate features, we take the average of all layers, which provides a compact representation of the entire track. Mathematically, for a song $\mathcal{S}$ with $T$ frames, the resulting final MERT feature vector is calculated as:
\begin{equation}
\mathcal{M}(\mathcal{S}) = \frac{1}{T}\sum_{t=1}^{T}\mathcal{M}_n(\mathcal{S}_t);   n \in \{1 ... N\}
\end{equation}
where, $\mathcal{M}_n$ is the MERT latent features at the $n$-th layer and $N$ is the total number of layers (25 in our case) in the MERT feature extractor. This results in a $N \times $1024-dimensional vector representation of the musical content for each song, i.e., $\mathcal{M}(\mathcal{S}) \in \mathbb{R}^{N \times 1024}$.

For the metadata features, as the dataset contains partially available metadata, we first encode the existing metadata fields using the Google flan-t5 model~\footnote{\url{https://huggingface.co/google/flan-t5-small}}. This sentence transformer encodes the metadata fields into a 768-dimensional vector. Mathematically, for a song $\mathcal{S}$ with metadata $\mathcal{T}$, the encoded metadata feature vector is: $\mathcal{F}(\mathcal{S})$, with $\mathcal{F} \in \mathbb{R}^{768}$.

To retrieve the most similar songs to a given song $S$, we concatenate the musical and metadata feature vectors and compute the cosine similarity between the query song and all other songs in the dataset. Since the dimensions of the MERT audio features and the metadata features are different, we align the MERT features using sparse random projection -- a dimensionality reduction technique to match the 768 dimensions of the metadata embeddings. Mathematically,  
\begin{equation}
\mathcal{X}(\mathcal{S}) = \lambda_1 \times (\mathcal{P}(\mathcal{M(\mathcal{S})}) + \lambda_2 \times \mathcal{F(\mathcal{S})}
\label{eq:ret}
\end{equation}
Where $\mathcal{P}(.)$ is the sparse random projection function, $\lambda_1$ and $\lambda_2$ are hyperparameters to balance the contribution of audio and metadata features such that $\lambda_1 + \lambda_2 = 1$. Here, $\mathcal{X} \in \mathbb{R}^{768}$. The similarity score between a query song $\mathcal{Q}$ and a song $\mathcal{S}$ in the dataset is calculated as:
\begin{equation}
\text{Similarity}(\mathcal{Q},\mathcal{S}) = \frac{\mathcal{X}(\mathcal{Q})^\top \mathcal{X}(\mathcal{S})}{||\mathcal{X}(\mathcal{Q})|| \times ||\mathcal{X}(\mathcal{S})||}
\end{equation}
Using this retrieval system, we can identify songs that are similar in both musical and metadata features to the query song. For each song in our dataset, we retrieve the top 10 most similar songs based on this similarity metric. We use the (partial) metadata and captions generated from the audio of these similar songs as the in-context examples for the large language model to impute the missing metadata fields, as described in the next section.

\subsection{Metadata Imputation}
To impute the missing metadata fields, we employ the capabilities of large language models. We have opted for a locally hosted large language model to maintain full control over the system and avoid dependence on remote servers. Specifically, we utilize Llama-2, a large language model with 7 billion parameters~\footnote{\url{https://huggingface.co/meta-llama/Llama-2-7b-hf}}, which is hosted on our own infrastructure.
\begin{figure}[t]
    \centering
    \includegraphics[width=0.5\textwidth]{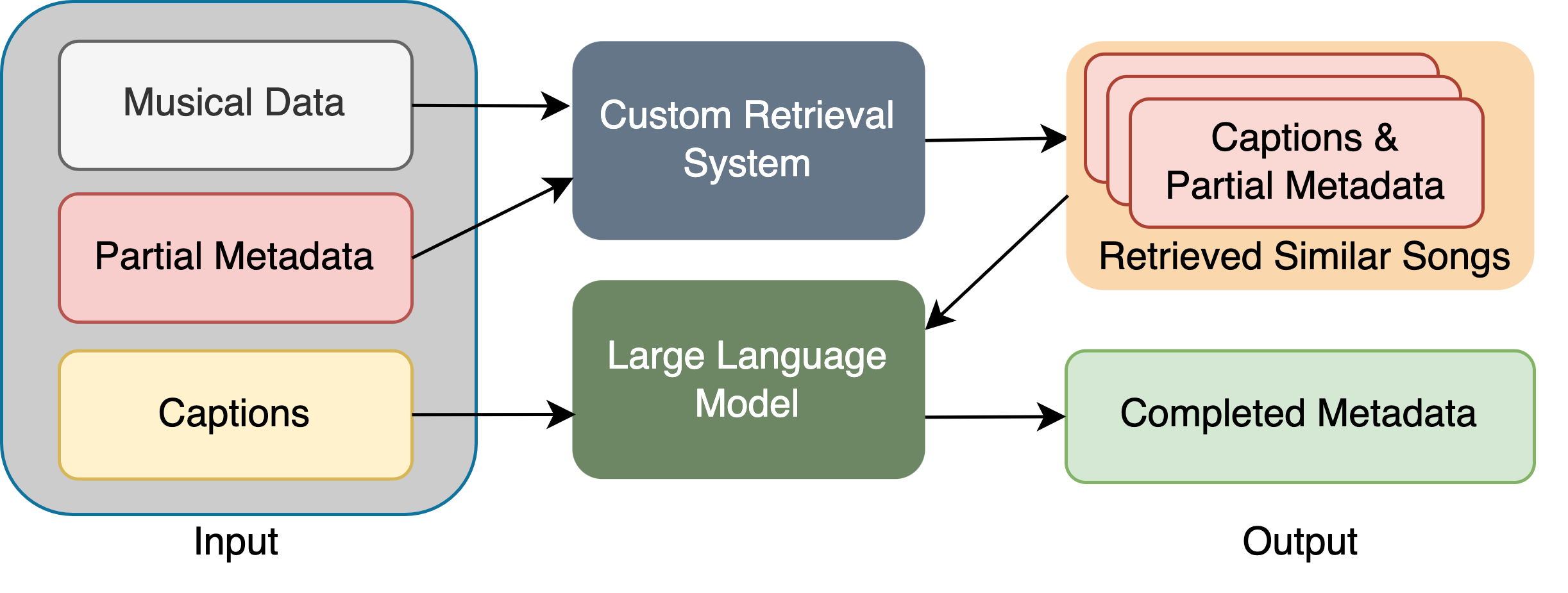}
    \caption{The metadata imputation process.}
    \label{fig:metadata_imputation}
    \vspace{-5pt}
\end{figure}
Algorithm~\ref{al:algo1} shows the pipeline for metadata imputation. For each song in the \textbf{\dataset}, we first retrieve the top 10 most similar songs using the retrieval system described earlier. We then construct an in-context examples by concatenating the metadata and captions of these top 10 similar songs. In-context learning has proven to be highly effective in empowering large language models to generate meaningful and relevant outputs~\cite{dong2022survey}. We leverage this approach by prompting the language model with these in-context examples and provide the caption of a target song, asking the large language model to infer the corresponding metadata~\footnote{\url{https://github.com/AMAAI-Lab/JamendoMaxCaps}}. This process allows the language model to draw insights from the contextual information of similar songs to impute the missing metadata fields for the target song.


\begin{algorithm}[h]
\caption{Imputing Missing Metadata}
\label{al:algo1}
\begin{algorithmic}[1]
\FOR{each song $\mathcal{Q} \in \text{\textbf{\dataset}}$}
    \STATE \textbf{Step 1: Retrieve Similar Songs} \\
    Retrieve the top $k$ similar songs, denoted as:
    \[
    \{\mathcal{S}_1, \mathcal{S}_2, \dots, \mathcal{S}_k\}
    \]
    using the retrieval system.
    \STATE \textbf{Step 2: Construct In-Context Example} \\
    Formulate the in-context example $\mathcal{E}$ as:
    \[
    \mathcal{E} = \Big((\mathcal{M}(\mathcal{S}_i), \mathcal{C}(\mathcal{S}_i)) \;\big|\; \mathcal{S}_i \in \{\mathcal{S}_1, \mathcal{S}_2, \dots, \mathcal{S}_k\}\Big)
    \]
    \textbf{Note:} $\mathcal{M}(\mathcal{S}_i)$ -- metadata; $\mathcal{C}(\mathcal{S}_i)$ -- caption of song $\mathcal{S}_i$.
    \STATE \textbf{Step 3: Prompt Language Model} \\
    Use the language model $\mathcal{L}$ with $\mathcal{E}$ and the caption of the target song $\mathcal{Q}$, denoted as $\mathcal{C}(\mathcal{Q})$:
    \[
    \mathcal{M}(\mathcal{Q}) = \mathcal{L}(\mathcal{E}, \mathcal{C}(\mathcal{Q}))
    \]
    \STATE \textbf{Step 4: Infer Metadata} \\
    Infer the missing metadata fields $\mathcal{M}(\mathcal{Q})$ for song $\mathcal{Q}$.
\ENDFOR
\end{algorithmic}
\end{algorithm}

\section{Evaluation And Statistics}
Here, we first discuss the dataset creation process, and then detail the results of the metadata imputation.

\subsection{\textbf{\dataset} Dataset}
To create the dataset, we first collected audio tracks and associated metadata from the Jamendo music platform, a widely used online repository of Creative Commons-licensed music. We used their publicly accessible API to download instrumental songs that were uploaded to the platform between 2008 and 2023 (both years included), downloading the raw audio tracks and extracting the corresponding metadata fields. This process yielded a total of 362,649 audio tracks with partial metadata. We then cleaned the dataset by removing tracks with very short durations, resulting in a final set of 362,238 tracks with an average duration of 4 minutes and 12 seconds. A brief overview of some of the properties of the dataset is presented in Table~\ref{tab:dataset_summary}.

Next, we use the \textit{Qwen2-Audio}~\cite{chu2024qwen2} to generate a caption that describes the musical content of each track in natural language. We take the caption of the first 30 seconds(~\ref{sec:cap}) for computational efficiency. This allows us to pair each audio track with a descriptive caption.

Finally, we apply the metadata imputation approach discussed in the previous section. We observe that since all songs in our dataset are instrumental, \textbf{`lang'} (language) and \textbf{`gender'} (gender of singer) are always empty. We discard these for the evaluation of the imputation task. Considering only the rest of the metadata fields, 261,935 songs had partial metadata, while the remaining 100,303 tracks had complete metadata information. We focus on imputing the missing metadata fields for the songs with partial metadata. To create the final dataset, we concatenated the songs, their corresponding captions, and the (imputed where applicable) metadata information. A couple of examples of metadata imputation are given in Table~\ref{tab:metadata_example}. 

\begin{table}[ht]
    \centering
    \sisetup{table-number-alignment = center, round-mode = places, round-precision = 3}
    \caption{Statistical Summary of the Dataset}
    \label{tab:dataset_summary}
    \begin{tabular}{l S[table-format=6.2] S[table-format=6.2] S[table-format=6.2,input-symbols = {-}]}
        \toprule
        \multirow{2}{*}{\textbf{Measure}} & \multicolumn{3}{c}{\textbf{Statistics}} \\
        \cmidrule(lr){2-4}
        & \textbf{Mean} & \textbf{Median} & \textbf{Mode} \\
        \midrule
        Length (seconds) $^{\mathrm{a}}$ & 252.30 & 209.853 & 124.959 \\
        SNR (dB) $^{\mathrm{b}}$       & 0.03 & 0.001 & -0.01 \\
        RMS Energy $^{\mathrm{b}}$      & 0.15  & 0.139  & 0.117  \\
        Zero Crossing Rate $^{\mathrm{b}}$ & 0.047  & 0.041  & 0.029  \\
        \bottomrule
    \end{tabular}
    \raggedright \textit{ $^{\mathrm{a}}$Dataset has 362,238 examples.\\ $^{\mathrm{b}}$SNR = Signal-to-noise ratio; RMS Energy = Signal's overall power; \\Zero Crossing Rate = How often a signal crosses the zero amplitude level, indicating the rate of sign changes.}
\end{table}

\begin{table*}[ht]
    \centering
    \caption{Comparison of Metadata Before and After Imputation$^{\mathrm{a}}$} 
    \label{tab:metadata_example}
    \renewcommand{\arraystretch}{1.5}  
    \begin{tabular}{p{5 cm} p{6 cm}}  
        \toprule
        \textbf{Original Metadata} & \textbf{Imputed Metadata} \\
        \midrule
        \begin{minipage}[t]{\linewidth} 
        {\textbf{`vocalinstrumental'}: `instrumental',\\
        \textbf{`lang'}: `',\\
        \textbf{`gender'}: `',\\
        \textbf{`speed'}: `medium',\\
        \textbf{`tags'}: {\textbf{`genres'}: [], \textbf{`instruments'}: [], \\
        \textbf{`vartags'}: []} }
        \end{minipage}
        &
       \begin{minipage}[t]{\linewidth}
        {\textbf{`vocalinstrumental'}: `instrumental',\\
        \textbf{`lang'}: `',\\
        \textbf{`gender'}: `',\\
        \textbf{`speed'}: `medium',\\
        \textbf{`tags'}: {\textbf{`genres'}: [`\textcolor{added}{pop}', `\textcolor{added}{electronic}', `\textcolor{added}{hip-hop}'],\\
                 \textbf{`instruments'}: [`\textcolor{added}{bass}', `\textcolor{added}{drums}'],\\
                 \textbf{`vartags'}: [`\textcolor{added}{intense}', `\textcolor{added}{energetic}']}}
        \end{minipage}
        \\
        \hline
                \begin{minipage}[t]{\linewidth} 
        {\textbf{`vocalinstrumental'}: `instrumental',\\
        \textbf{`lang'}: '',\\
        \textbf{`gender'}: '',\\
        \textbf{`speed'}: 'medium',\\
        \textbf{`tags'}: {\textbf{`genres'}: [], \textbf{`instruments'}: [], \\
        \textbf{`vartags'}: []} }
        \end{minipage}
        &
       \begin{minipage}[t]{\linewidth}
        {\textbf{`vocalinstrumental'}: `instrumental',\\
        \textbf{`lang'}: '',\\
        \textbf{`gender'}: '',\\
        \textbf{`speed'}: 'medium',\\
        \textbf{`tags'}: {\textbf{`genres'}: [`\textcolor{added}{cyberpunk}'],\\
                 \textbf{`instruments'}: [],\\
                 \textbf{`vartags'}: [`\textcolor{added}{synthwave}']}}
        \end{minipage}
        \\
        \hline
        \multicolumn{2}{p{11 cm}}{$^{\mathrm{a}}$
        \textit{\textbf{Bold keys} represent the metadata fields. \textcolor{added}{Green text} highlights newly imputed values.} 
        }
        \\
    \end{tabular}
\end{table*}

\begin{figure*}[ht]
  \centering
  
  \includegraphics[width=\textwidth]{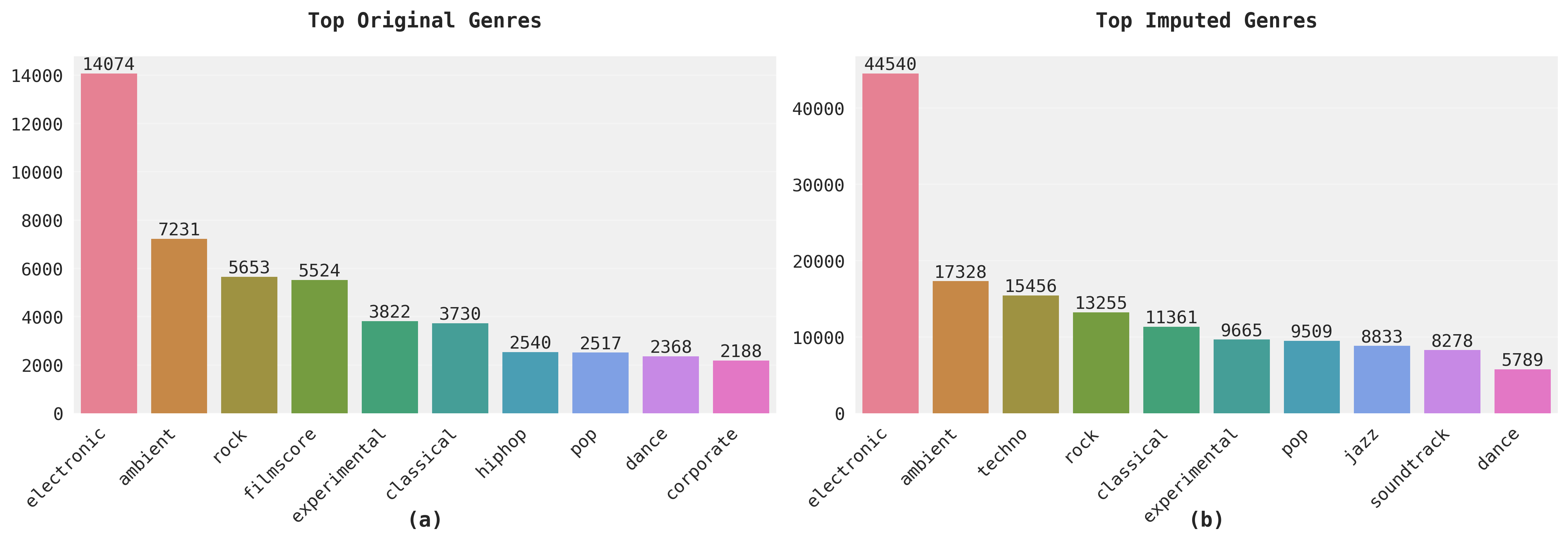}
  \caption{Distribution of \textbf{genres} in original and imputed metadata.}
  \label{fig:genre}
\end{figure*}
\begin{figure*}[ht]
  \centering
  \includegraphics[width=\textwidth]{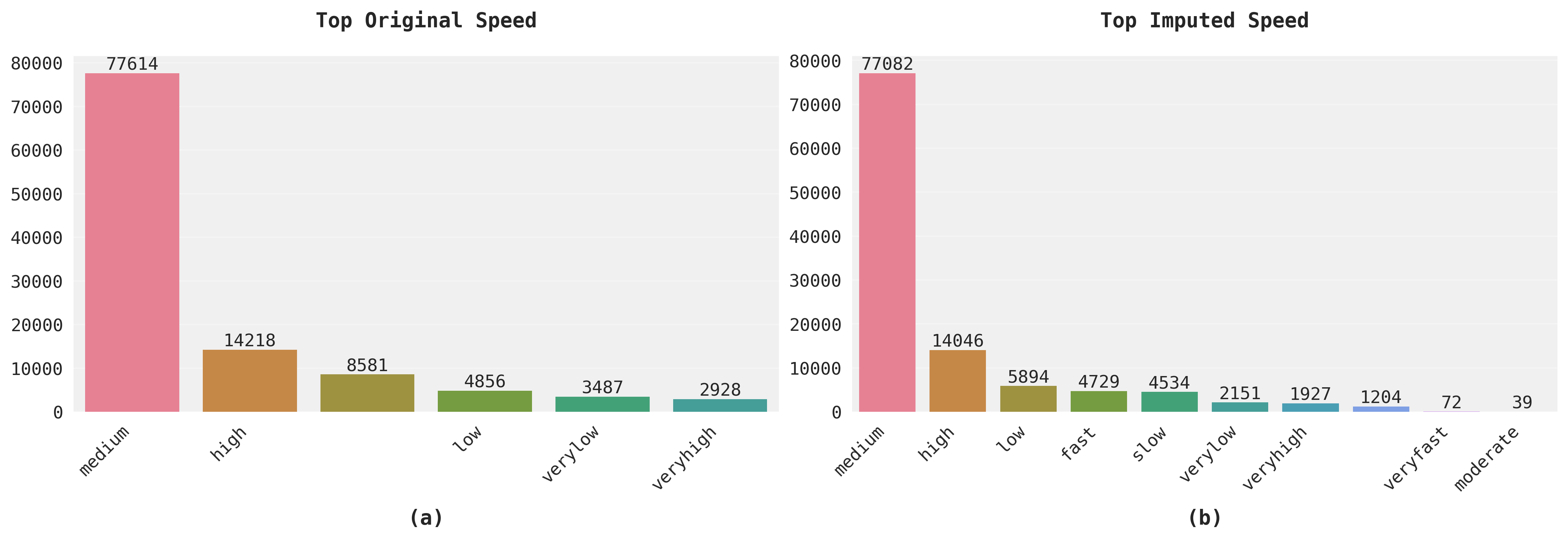}
  \caption{Distribution of \textbf{speed} in original and imputed metadata.}
  \label{fig:speed}
\end{figure*}
\begin{figure*}[ht]
  \centering
  \includegraphics[width=\textwidth]{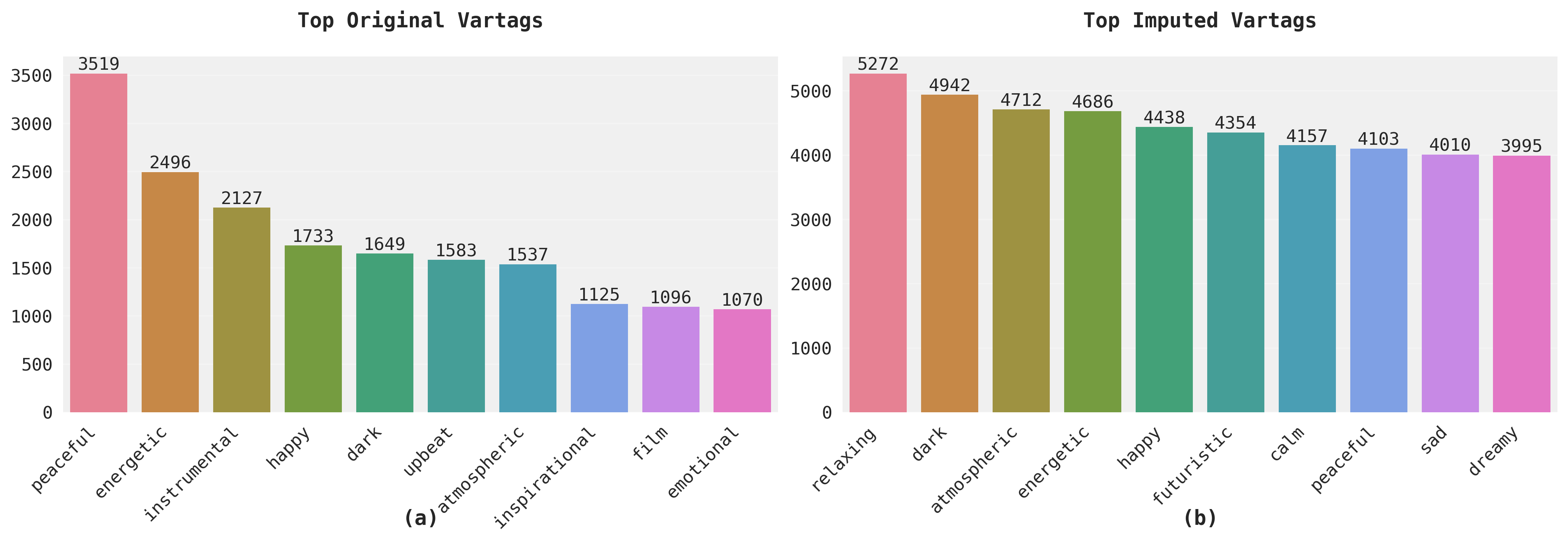}
  \caption{Distribution of \textbf{variable tags} in original and imputed metadata.}
  \label{fig:vartags}
\end{figure*}

Furthermore, we provide a summary of the distribution of some key metadata fields in the dataset to offer additional insight into its composition. This includes an analysis of the distribution of genres, and speed characteristics of the songs, which can help researchers and practitioners better understand the dataset and its potential applications. Figure~\ref{fig:genre} shows that the metadata imputation process adds more examples of previously underrepresented categories, such as ``Techno" and ``Jazz," while also retaining the representation of major genres ``Electronic" and ``Rock". This imputation effectively fills in the missing metadata fields, resulting in a more comprehensive and informative dataset representation. Moreover, the prevalence of ``Electronic" and ``Rock" genres in the dataset suggests the inclusion of more contemporary musical works, which aligns with the time period of our data collection efforts (2008-2023). In the case of \textbf{speed}, the imputation process enhances the granularity of the metadata by introducing additional categories, including ``Fast", ``Slow", ``Very Fast", and ``Moderate", which were not initially present significantly in the original data. While ``Medium" remains the predominant category, the imputed metadata offers a more nuanced distribution, capturing a broader range of tempo variations. Noteworthy, in the original data, missing/nothing was the third most common category, but the imputation process successfully eliminated this gap, adding valuable information to the dataset. Figure~\ref{fig:vartags} illustrates a significant enhancement in the diversity and coverage of \textbf{vartags}, providing more representation of musical characteristics. While the original data concentrated on a few prevalent tags such as ``peaceful" and ``energetic," the imputed metadata introduces a wider range of descriptors, including ``relaxing," ``futuristic," and ``dreamy", which better encapsulate the emotional and atmospheric elements of the music.

In summary, the \textbf{\dataset}  dataset, created through a combination of data collection, caption generation, and metadata imputation, offers a comprehensive and diverse music repository with rich descriptive information.

\begin{table}[ht]
    \centering
    \caption{Subjective evaluation results for different retrieval configurations }
    \label{tab:hyperparam}
    \begin{tabular}{ccc}
        \toprule
        \textbf{Retrieval Configuration} & \textbf{Eval 1} & \textbf{Eval 2} \\
        \midrule
        $\lambda_1 = 0$; $\lambda_2 = 1$ (Only metadata) & 2.0 & 3.0 \\
        $\lambda_1 = 1$; $\lambda_2 = 0$ (Only musical features) & 3.0 & 3.0 \\
        $\lambda_1 = 0.4$; $\lambda_2 = 0.6$ (Metadata favored) & 4.0 & 4.0 \\
        $\lambda_1 = 0.6$; $\lambda_2 = 0.4$ (Musical features favored) & \textbf{6.0} & \textbf{6.0} \\
        \bottomrule
    \end{tabular}
    \raggedright \textit{The ratings represent the scores given by two independent participants on a 7-point Likert scale~\cite{likert1932technique} -- from 1 (very bad) to 7 (very good).}
    \label{tab:retrieval_evaluation}
\end{table}
\begin{table*}[ht]
\centering
\caption{Evaluation of Metadata Imputation with Generic Examples vs Retrieval Examples}
\begin{tabular}{l|ccc||ccc}
\toprule
\textbf{Evaluation Metric} & \multicolumn{3}{c||}{\textbf{ Imputation with Generic Examples}} & \multicolumn{3}{c}{\textbf{Imputation with Retrieval Examples}} \\
\midrule
 & \textbf{Genres} & \textbf{Speed} & \textbf{Vartags} &\textbf{Genres} & \textbf{Speed} & \textbf{Vartags} \\
\midrule
\textbf{BERT-Score $\uparrow$} & 0.51 & 0.70 & 0.35 & \textbf{0.58} & \textbf{0.73} & \textbf{0.38} \\
\textbf{BLEU-1 $\uparrow$} & 0.17 & 0.61 & 0.02 & \textbf{0.19} & \textbf{0.64} & \textbf{0.03} \\
\textbf{BLEU-2 $\uparrow$}  & 0.07 & 0.58 & 0.01 & \textbf{0.08} & \textbf{0.60} & 0.01 \\
\textbf{BLEU-3 $\uparrow$} & 0.05 & 0.57 & 0.01 & \textbf{0.06} & \textbf{0.60} & \textbf{0.02} \\
\textbf{BLEU-4 $\uparrow$} & 0.05 & 0.57 & 0.01 & 0.05 & \textbf{0.60} & 0.01 \\
\bottomrule
\end{tabular}

\label{tab:metadata_imputation_final}
\end{table*}

\subsection{Retrieval System Settings}
Since Qwen2-Audio~\cite{chu2024qwen2} is an established model, we do not evaluate its performance. Instead, we focus on the strength of the retrieval system and the metadata imputation approach. 

To evaluate the quality of the retrieval system used for metadata imputation, we explore different combinations of musical features and metadata features as outlined in Equation~\ref{eq:ret}. We vary the values of $\lambda_1$ and $\lambda_2$ in four distinct ways: 
\begin{itemize}
    \item $\lambda_1 = 0$; $\lambda_2 = 1$ -- Only rely on metadata information 
    \item $\lambda_1 = 1$; $\lambda_2 = 0$ -- Only rely on musical features 
    \item $\lambda_1 = 0.4$; $\lambda_2 = 0.6$ -- Favour metadata features more over musical features 
    \item  $\lambda_1 = 0.6$; $\lambda_2 = 0.4$ -- Favour musical features more over metadata features
\end{itemize}

We conduct a small-scale subjective evaluation involving two independent evaluators to assess the relevance and similarity of the retrieved songs. The participants were asked to rate the overall quality of the retrieval on a 7-point Likert scale~\cite{likert1932technique} -- 1 being least similar, 7 being most similar songs. From the dataset, we select five songs and retrieve similar songs based on the four variations of $\lambda_1$ and $\lambda_2$ mentioned above. The results are presented in Table~\ref{tab:hyperparam}, which shows that the combination of $\lambda_1=0.6$ and $\lambda_2=0.4$ outperformed the other three options. Looking at the stark difference in the performance, we decided not to carry out any further subjective evaluation (with more participants) related to $\lambda$. 

\subsection{Objective Evaluation}
To ensure the reliability of our metadata imputation approach and to verify that the local language model we use does not hallucinate, we conduct two-fold objective evaluations. 

To evaluate the accuracy of the metadata imputation, we first identify the examples that have valid entries in the original metadata for the `genres', `speed', and `vartags' fields. We then generate imputed values for these fields using our approach and compare them to the original values for 5,000 examples per field. As the primary metric, we employ BERT-Score~\cite{zhang2019bertscore} to measure the semantic similarity between the original and imputed values. Since the imputation can generate multiple predictions, this allows us to establish a semantic similarity score between the original and imputed data. Additionally, we utilize n-gram metrics such as BLEU1 to BLEU4 (B1, B2, B3, B4)~\cite{papineni2002bleu} to assess the exact overlap between the original and imputed values. Please note that the primary purpose of BLEU scores is to compare machine translations with ground truth, rather than evaluating semantic similarity.

Next, to evaluate the effectiveness of in-context learning for metadata imputation, we generate a secondary baseline that learns from a set of 10 \textit{random} examples during the in-context learning phase, rather than utilizing the most similar songs retrieved through our retrieval system. This approach allows us to assess the impact of leveraging contextually relevant information versus a generic set of examples when imputing missing metadata fields. We also compute the BERT-Score and BLEU metrics (BLEU1 through BLEU4) to evaluate the semantic similarity and n-gram overlap between the original and imputed metadata values, respectively. This evaluation is on the same 5,000 examples per metadata field, as discussed before. The results are compiled in Table~\ref{tab:metadata_imputation_final}. Results demonstrate that metadata imputation using retrieval examples consistently outperforms imputation with generic examples across all evaluation metrics. Specifically, the BERT-Score is higher for all fields when retrieval examples are used, indicating a better semantic alignment between the original and imputed metadata. The relatively high BERT-Scores (e.g., 0.58 for genres, 0.73 for speed, when using retrieval examples) suggest that the imputation method is producing meaningful values that closely resemble the original metadata. On the other hand, the BLEU scores, especially for genres and vartags, are quite low, which reflects the nature of the imputation task. These low scores (e.g., 0.05 for genres and 0.01 for vartags) highlight the challenge of achieving perfect overlap with the original values, as metadata fields like genres and vartags often contain a broad range of possible terms, making exact matches unlikely. In contrast, the higher BLEU scores for speed (e.g., 0.64 for retrieval examples) can be attributed to the limited and more constrained set of values (e.g., `fast', `slow'), which results in a higher likelihood of overlap. Overall, these results confirm that our metadata imputation approach, which leverages contextually relevant retrieval examples, is effective in generating meaningful and accurate imputed values. 
\begin{table}[ht]
\centering
\renewcommand{\arraystretch}{1.2}
\caption{Results of the listening study evaluating imputed metadata}
\label{tab:listening_study}
\begin{tabular}{lc}
\toprule
\textbf{Criterion} & \textbf{Mean\(\pm\)SD} \\
\midrule
Human-like Generation & 5.32\(\pm\)0.31 \\
Genres Matching       & 5.09\(\pm\)0.69 \\
Speed Matching        & 5.82\(\pm\)0.43 \\
Vartags Matching      & 5.59\(\pm\)0.45 \\
\bottomrule
\end{tabular}
\end{table}
\subsection{Subjective Evaluation}
To subjectively verify the quality of the imputation, we conduct a listening study using the PsyToolkit platform\footnote{\url{https://www.psytoolkit.org/}}~\cite{stoet2010psytoolkit}. Study participants are asked to listen to 10 randomly selected audio files and rate the associated imputed metadata on four criteria: 1) the extent to which the metadata could have been generated by a human, 2) the accuracy of the imputed \textbf{genres}, 3) the accuracy of the imputed \textbf{speed}, and 4) the accuracy of the imputed \textbf{vartags}. A 7-point Likert scale~\cite{likert1932technique} is used for scoring, with 1 being the lowest and 7 being the highest. A total of 15 participants took part in the listening study.
~\footnote{Questions and samples: \url{https://github.com/AMAAI-Lab/JamendoMaxCaps}}.

The results of the listening study are presented in Table~\ref{tab:listening_study}. We achieve consistently high scores, with all metrics rated above 5.09 out of 7, indicating a strong overall performance.  \textbf{`speed'} and \textbf{`vartags'} matching received particularly impressive scores of 5.82 and 5.59 respectively, with low standard deviations (0.43, 0.45) suggesting reliable and consistent performance across different musical pieces. The high score for human-like generation (5.32\(\pm\)0.31) indicates that our system successfully captures the characteristics that human annotators would typically provide. Even the relatively lower score for \textbf{`genres'} matching (5.09\(\pm\)0.69) remains above the median point of the scale, demonstrating competent genre classification capabilities. These results suggest that our approach could serve as a reliable tool for automatic metadata generation.

\section{Discussion}

While our evaluation confirms the effectiveness of the \textbf{\dataset} dataset in improving metadata completeness and retrieval quality, several challenges and areas for future refinement remain. One primary challenges is the reliance on imputed metadata, which, despite leveraging a local large language model (LLLM) and retrieval-based contextualization, remains inherently constrained by the quality and representativeness of the retrieved examples. As demonstrated in our evaluation, the imputed metadata fields, particularly those related to genre and variable tags (e.g., mood descriptors), exhibit lower BLEU scores, suggesting a notable degree of variability in how different musical elements are categorized. This variability arises from the inherent subjectivity in musical taxonomy, where the same piece may be classified under different labels depending on context and personal interpretation. Our approach does not entirely eliminate inconsistencies, especially for songs with highly unique stylistic elements that may not align well with the retrieved examples.  

Another limitation comes from potential biases in the retrieval system. The dataset composition influences the quality of metadata imputation, as genres with a higher representation, such as ``Electronic" and ``Rock", benefit from a richer pool of similar examples, whereas underrepresented genres, such as ``Jazz" or ``Experimental", may suffer from less reliable imputations due to the scarcity of relevant contextual data. This imbalance can lead to an over-representation of dominant categories in the imputed metadata, potentially reinforcing dataset biases rather than mitigating them. A more sophisticated genre-aware retrieval strategy can be explored to ensure a fairer distribution of metadata across all musical styles.  


The choice of a locally hosted LLM, while advantageous in terms of cost, privacy, and control, may also present certain trade-offs. Although our model demonstrates strong performance in metadata generation, it may lack the extensive pretraining and fine-tuning capabilities of larger cloud-based models, which are often trained on more diverse and extensive datasets. A promising direction for future work is to explore hybrid approaches that integrate local inference with selective refinements using advanced cloud-based models, ensuring both efficiency and high-quality metadata generation.  

\section{Conclusion}
The \textbf{\dataset} dataset represents a significant step forward in enhancing music-language understanding. By addressing metadata sparsity and providing a large-scale, freely accessible dataset, this work enables more effective research in music information retrieval, representation learning, and generation.

\section*{Acknowledgment}
This work has received support from SUTD's Kickstart Initiative under grant number SKI 2021\_04\_06 and MOE under grant number MOE-T2EP20124-0014. We acknowledge the use of ChatGPT for grammar refinement and paraphrasing.

\bibliographystyle{abbrv}
\bibliography{references.bib}

\end{document}